\documentclass{article}
\usepackage{hyperref}
\usepackage{amsmath}
\usepackage{graphicx}
\usepackage{amsfonts}
\usepackage{amssymb}
\usepackage[english]{babel}
\begin{document}

\title{Elements of Fedosov Geometry in Lagrangian BRST Quantization}
\author{\textsc{A.A.
Reshetnyak}\thanks{E-mail addresses: A-Reshetnyak@yandex.ru, reshet@tspu.edu.ru}\\
Tomsk State Pedagogical University, 634041 Tomsk, Russia \\}
\date{}
\maketitle

\begin{abstract}
A Lagrangian BRST quantization for generic gauge
 theories in general irreducible non-Abelian hypergauges  is proposed
 on a  basis of the multilevel Batalin--Tyutin formalism and a special
 BV--BFV dual description for a reducible gauge model in a symplectic
 supermanifold $\mathcal{M}_0$ locally parameterized by antifields
 for Lagrangian multipliers and by the fields of the BV method. The
quantization rules are based on a set of nilpotent anticommuting
operators $\Delta^\mathcal{M}, {V}^\mathcal{M}, {U}^\mathcal{M}$
defined using some odd and even symplectic structures in a supersymplectic
manifold $\mathcal{M}$ whose local representation is an odd (co)tangent
bundle over $\mathcal{M}_0$ provided by the choice of a flat Fedosov
connection and a compatible non-symplectic metric in $\mathcal{M}_0$.
The generating functional of Green's functions is constructed in terms
of general coordinates in $\mathcal{M}$ with the help of contracting
homotopy operators with respect to ${V}^\mathcal{M}$ and ${U}^\mathcal{M}$.
We prove the gauge independence of the S-matrix and derive the Ward identity.
\end{abstract}


\paragraph{1. Introduction} The conventional form of Lagrangian
\cite{BV1} (\cite{BLT1}
) and Hamiltonian \cite{BFVH} (\cite{BLT2,Henn1}) quantization
for general gauge theories, implementing the BRST (BRSTantiBRST)
symmetry and maintaining locality and global symmetries, was developed
around 15--20 years ago and is largely sufficient for perturbative
quantization of gauge models set within the variational principle.
Nonetheless, it is of interest to study some additional problems related
to geometrically covariant descriptions of the quantization procedure
reflecting the global and invariant properties of a manifold of field
variables in specific models. This activity was originated by the Lagrangian
multilevel formalism \cite{BatalinTyutin1} and was continued by the Hamiltonian
coordinate-free formalism \cite{BT0}, based on the Weyl symbols, as
well as by the (modified) triplectic \cite{BMS} (\cite{GGL}) BRSTantiBRST scheme.
On the other hand, it is closely related (via the notion of supertime $\chi=(t,\theta)$)
with the problem of an equitable description for the dynamics and
BRST transformations of a model when treated by the Lagrangian
\cite{R0,R1,GMR,LMR} and Hamiltonian \cite{BBD1,GrigorievDamgaard,BBD2}
superfield quantization.

A solution of the first series of problems with reference to the general aspects
of quantization is the construction of a $\star$-product by Kontsevich \cite{Kontsevich}
within a deformation quantization for the algebra of functions in an arbitrary
Poisson manifold $\mathcal{M}_{\mathrm{P}}$ to which one applies \cite{CattaneoFelder}
a topological Poisson $\sigma$-model defined in $\mathcal{M}_{\mathrm{P}}$, whose
field-antifield spectrum, following the AKSZ approach
\cite{AlexandrovKontsevichSchwarzZaboronsky} in an $N=2$
superfield (non-spacetime) formulation, coincides with the corresponding fields
and antifields of the BV method. One of these problems is a construction of deformation
quantization for dynamical systems with second-class constraints and symplectic manifolds
\cite{BLG} on a basis of the BFV--BRST \cite{BFVH} conversion methods \cite{FaShBaFrBaTy},
as well as for non(-Lagrangian)-Hamiltonian gauge theories \cite{SharLyahov}, such as higher-spin
gauge fields \cite{Vasiliev}, where one essentially uses a symmetric connection compatible
with a symplectic structure, i.e., the Fedosov connection \cite{Fedosov,GRSh}.

While the Lagrangian BRSTantiBRST (superfield) quantization \cite{GLN} (\cite{GGLM}) defined
in general coordinates maintains a tensor character of compatible differential operations, i.e.,
extended antibrackets and odd operators $(\Delta^a, \mathcal{V}^a,\mathcal{U}^a), a=1,2$,
only in the case of a \emph{flat} Fedosov connection in a supermanifold $\widetilde{\mathcal{M}}_0$
using Darboux coordinates parameterized by the fields $\phi^A$ of the BV
method and by the corresponding antifields $\overline{\phi}_A$, being
sources for the commutator of BRSTantiBRST transformations
\cite{BLT1}
, the construction of Lagrangian BRST quantization in irreducible
\cite{BatalinTyutin1} and reducible (originally introduced in
\cite{R1}) non-Abelian hypergauges\footnote{The study of the influence
of \emph{reducibility} and \emph{non-Abelian properties} of
hypergauges on specific gauge models in quantum calculations
presents an essential part of \cite{MoshinReshetnyak}.},
in fact, does not utilize\footnote{Except for a special
connection $F_p(\Gamma) = \delta
\{\ln(\rho[\Gamma])\}/\delta \Gamma^p$ in \cite{BatalinTyutin1},
expressed using a density functional $\rho[\Gamma]$ in the
``first-level'' supermanifold $\mathcal{N} = \{\Gamma^p\}$
determining the (anti)fields ($\phi^*_A$) $\phi^A$ of the BV
method in the Darboux coordinates $\Gamma^p = (\phi^A,\phi^*_A)$.}
the concept of Fedosov connection. The local superfield BRST
quantization \cite{R1,GMR} shows that restricting the consideration
to the ingredients of the \emph{first-level} formalism, in view of
a special nature of Lagrangian multipliers $\lambda^a$ for hypergauges
$G_a(\Gamma)$, is insufficient to introduce a covariant derivative
in $\mathcal{N}$. For this purpose, one must utilize not only
$\lambda^a$, but also the antifields $\lambda^*_a$ that arise
in the \emph{second-level} formalism \cite{BatalinTyutin1}.

This report aims essentially at the following:

\noindent
\begin{enumerate}
    \item Description of a gauge algebra for a reducible gauge
    model by means of a \emph{special BV--BFV duality} between odd
    $\mathcal{N}_{\mathrm{min}}$ and
    even $\mathcal{M}_{0{}\mathrm{min}}$  symplectic manifolds\footnote{In what
    follows, we often omit the prefix ``super'' in ``supermanifold''.}
    underlying the quantization procedure and intersecting with a manifold
    parameterized by the minimal-sector fields of the BV method.\vspace{-1ex}
    \item Investigation of a supersymplectic structure of
    the quantization manifold\footnote{$\mathcal{M}$ locally
    represents a vector bundle over the manifold $\mathcal{M}_0$,
    $\mathcal{M} \rightarrow \mathcal{M}_0$, so that
    $\mathcal{M}_0 \supset \mathcal{M}_{0{}\mathrm{min}}; \mathcal{N} \supset
    \mathcal{N}_{\mathrm{min}}; \mathcal{M}_0, \mathcal{N} \subset
    \mathcal{M}$.}
    $\mathcal{M}$ compatible with the requirement of
    anticommutativity for a set of nilpotent operators $\Delta^{\mathcal{M}},
    {V}^{\mathcal{M}}, {U}^{\mathcal{M}}$.\vspace{-1ex}
    \item Formulation of quantization rules for gauge models using general coordinates
    in $\mathcal{M}$ with an essential use of operators $\mathcal{V}^{\ast{}},
    \mathcal{U}^{\ast{}}$ (constructed in terms of even and odd Poisson brackets),
    whose difference $(\mathcal{U}^{\ast{}}-\mathcal{V}^{\ast{}})$ is a contracting
    homotopy with respect to ${V}^{\mathcal{M}}$ for an operator $N^{\mathcal{M}}$
    nondegenerate for nonconstant elements of $C^{\infty}(\mathcal{M})$.
\end{enumerate}

\paragraph{2. Special BV--BFV dual description for a gauge model}
 Recall that an $L$-stage reducible gauge theory based on the variational
 principle for classical fields $A^i$, $i=1,...,n=n_++n_- $ (with Grassmann
 parities $\varepsilon(A^i)=\varepsilon_{i}$ in condensed
 notation) is defined by a classical bosonic action
 $\mathcal{S}_0(A)$: $C^{\infty}\left(\mathcal{M}_{\mathrm{cl}}\right)
 \rightarrow \mathbb{R}$, $\mathcal{M}_{\mathrm{cl}} = \{A^i\}$,
 invariant with respect to gauge transformations, $\delta A^i$ =
 $R^i_{\alpha_0}(A)\xi^{\alpha_0}$, $\alpha_0 = 1,...,m_0 = m_{0+} +
 m_{0-}$, $\varepsilon(\xi^{\alpha_0}) = \varepsilon_{\alpha_0}$,
 \begin{equation}
\mathcal{S}_0,_i R^i_{\alpha_0}(A) = 0, \; \mathrm{for}\;
\mathrm{rank}\left\|\mathcal{S}_0,_{ij}(A)\right\|_{\mathcal{S}_0,_k
= 0} = \overline{n}-\overline{m}_{-1}\equiv
(n_+,n_-)-({m}_{-1+},{m}_{-1-}), \mathcal{S}_0,_i \equiv {\delta
\mathcal{S}_0}/{\delta  A^i},
\label{1}%
\end{equation}
 with reducibility relations in condensed notation,
 for $s =1,...,L$,
\begin{eqnarray}
& & {{Z}}_{{\alpha}_{s-1}%
}^{\alpha_{s-2}}(A){{Z}%
}_{\alpha_{s}}^{\alpha_{s-1}}(A)= \mathcal{S}_0,_j {L}_{\alpha_{s}%
}^{\alpha_{s-2}j}\left( {A}%
\right)  , \;\alpha_s = 1,...,m_s = m_{s+} + m_{s-},\nonumber\\
& &\hspace{-2em}
\overline{m}_{s-1}\hspace{-0.1em}>\hspace{-0.1em}\sum\nolimits_{k=0}^{s-1}(-1)^{k}\overline
{m}_{s-k-2}\hspace{-0.1em}=\hspace{-0.1em}\mathrm{rank}\left\|
{Z}_{\alpha_{s-1}}^{\alpha_{s-2}}\right\|  _{\mathcal{S}_0,_k =
0}\hspace{-0.2em},  \; \overline{m}_{L}\hspace{-0.1em}=\hspace{-0.1em}\sum\nolimits_{k=0}^{L}(-1)^{k}%
\overline{m}_{L-k-1}\hspace{-0.1em}=\hspace{-0.1em}\mathrm{rank}\left\|
{Z}_{\alpha_{L}}^{\alpha_{L-1}}\right\|  _{\mathcal{S}_0,_k
= 0}\hspace{-0.2em},\nonumber\\
& & {\varepsilon}({Z}_{\alpha_{s+1}%
}^{\alpha_{s}})={\varepsilon}_{\alpha_{s}}+{\varepsilon
}_{\alpha_{s+1}},\ {Z}_{\alpha_{0}%
}^{\alpha_{-1}}\equiv{R%
}_{\alpha_{0}}^{i},\ {L}_{\alpha_{1}}^{\alpha_{-1}j}\equiv {K}_{\alpha_{1}}%
^{ij} = -(-1)^{(\varepsilon
_{i}+1)(\varepsilon_{j}+1)}{K}_{\alpha_{1}}^{ji}. \label{2}%
\end{eqnarray}
Definitions (\ref{1}) and (\ref{2}) partially determine the first-order
structure relations and functions of the gauge algebra,
and are encoded, due to the corresponding Koszul--Tate complex
resolution \cite{HT}, when extended (following the BV method) to an
odd-Hamiltonian description of the model in $\Pi
T^{\ast}\mathcal{M}_{\mathrm{min}} = \{\Gamma^{p_k}_k =
(\phi^{A_k},\phi^{\ast}_{A_k})|\phi^{A_k} = (A^i, C^{\alpha_s},
s=0,...,L), A_k=1,...,n_k=n+\sum^L_{r=0}m_r;
k=\mathrm{min}\}$,\footnote{The coordinates in $\Pi
T^{\ast}\mathcal{M}_{k}$ possess the following distribution of
Grassmann parity and ghost number \cite{BV1}: ($\varepsilon,
\mathrm{gh})C^{\alpha_s}=({\varepsilon}_{\alpha_{s}}+s+1,s+1),
\mathrm{gh}(A^i)=0, \varepsilon(\phi^{A_k})=\varepsilon_{A_k},
(\varepsilon, \mathrm{gh})\phi^{\ast}_{A_k} =
(\varepsilon_{A_k}+1,-1-\mathrm{gh}(\phi^{A_k}))$.} by means of
a bosonic functional and a classical master-equation in the minimal
sector \cite{BV1},
 \begin{eqnarray}
   &{}& S_k(\Gamma_k) =  \mathcal{S}_0(A) + \sum\nolimits^L_{s=0}\left(C^{\ast}_{\alpha_{s-1}}
   {{Z}}_{\alpha_{s}}^{\alpha_{s-1}}(A) + o(\phi^{\ast}_{k})\right)C^{%
   \alpha_{s}} + o(C^{\alpha_{s}}),\ \, (\varepsilon,\mathrm{gh})S_k = \vec{0}, \nonumber \\
   &{}& \left(S_k, S_k\right)^k = \frac{\delta_r S_k}{\delta
   \Gamma^p_k}\omega^{pq}_k(\Gamma_k)\frac{\delta_l S_k}{\delta
   \Gamma^q_k} = 0,\ \,
   \left\|\omega^{pq}_k\right\|=\mathrm{antidiag}\left(%
-\mathbf{{1}}_{n_k},\mathbf{{1}}_{n_k}\right). \label{3}
 \end{eqnarray}

The quantum action $S^{\psi}(\Gamma_k,\hbar)$ of the BV method (in
what follows, $k=\mathrm{ext}$) is constructed, first, by an extension
of $S_{\mathrm{min}}$, using the pyramids of ghosts and
Nakanishi--Lautrup fields, up to $S_k(\Gamma_k,\hbar)$
defined in $\Pi T^{\ast}\mathcal{M}_{k} = \{\Gamma^{p_k}_k =
(\phi^{A_k},\phi^{\ast}_{A_k})|\phi^{A_k} =
(\phi^{A_{\mathrm{min}}}, C^{\alpha_s}_{s'}, B^{\alpha_s}_{s'},
s'=0,...,s, s=0,...,L), A_k=1,...,n_k=n+\sum^L_{r=0}(2r+3)m_r;
k=\mathrm{ext}\}$, and, second, by imposing an Abelian hypergauge
corresponding to a phase anticanonical transformation in $\Pi
T^{\ast}\mathcal{M}_{k}$ for an $\hbar$-deformed $S_k(\Gamma_k,\hbar)$,
\begin{equation}
S_k(\Gamma_k)=S_{\mathrm{min}} + \sum\nolimits_{s=0}%
^{L}\sum\nolimits_{s^{\prime}=0}^{s}{C}_{s^{\prime}{}\alpha_{s}}^{\ast
}{B}_{s^{\prime}}^{\alpha_{s}};\
S^{\psi}(\Gamma_k,\hbar)=\exp\left[
\left(\psi(\phi_k),\ ^k\right)\right]%
S_{k}(\Gamma_{k},\hbar).\label{4}
\end{equation}
The functionals $[S_k,S^{\psi}](\Gamma_k,\hbar)$ obey a quantum
master-equation and provide its proper solutions in terms of a
nilpotent operator $\Delta^k$ constructed using a nondegenerate
antibracket in $\Pi T^{\ast}\mathcal{M}_{k}$, as well as using a trivial density
function $\rho(\Gamma_k)$, $\rho=1$, and $\omega^k_{pq}(\Gamma_k)$,
$\|\omega^k_{pq}\|$
= $\mathrm{antidiag}\left(%
\mathbf{1}_{n_k},-\mathbf{1}_{n_k}\right)$,
$\omega_k^{pq}\omega^k_{qr}=\delta^p_r$,
\begin{equation}\label{5}
    \Delta^{k}\exp\left\{  \frac{i}{\hbar}E(\Gamma_k,\hbar)\right\}
=0,\;E\in\{S^{\psi},S_{k}\},\
\Delta^{k}=\frac{1}{2}(-1)^{\varepsilon(\Gamma^{q})}\rho
^{-1}\omega^k_{qp}\left(  \Gamma^{p}_k,\rho\left( \Gamma ^{q}_k,\
\cdot\ \right)^k \right)^k.
\end{equation}
In the second-level formalism \cite{BatalinTyutin1}, the presence
of antifields $\lambda^\ast_a$ for Lagrangian multipliers
$\lambda^a$ introducing non-Abelian first-level hypergauges
$G_a(\Gamma_k)$ to the exponent of the path integral\footnote{In the case
of a fiber bundle $\Pi T^{\ast}\mathcal{M}_{k}$: $a=A_k, G_a =
G_{A_k}(\Gamma_k)=\left(\phi^{\ast}_{A_k}-\delta \psi /
\delta\phi^{A_k}\right)$ and
$\left[\lambda^a,\lambda^\ast_a\right] =
\left[\lambda^{A_k},\lambda^\ast_{A_k}\right]$, this corresponds
to the construction of $S^{\psi}(\Gamma_k,\hbar)$ in (\ref{4}).}
$Z^{(1)}$ allows one to construct a special BV--BFV dual description
in the cotangent bundle $T^{\ast}\mathcal{M}_k = \{x^{p_k}_k
= (\phi^{A_k}, \lambda^{\ast}_{A_k}),$
$(\varepsilon,\mathrm{gh})\lambda^{\ast}_{A_k}=(\varepsilon_{A_k},-2-\mathrm{gh}(\phi^{A_k}))\}$ for
a gauge theory of rank 1 [in general, in a sub-bundle
$\mathcal{N}_{\mathrm{aux}} \rightarrow \mathcal{M}_k$ of the
bundle $\Pi T^{\ast}(T^{\ast}\mathcal{M}_k)$, $\Pi
T^{\ast}(T^{\ast}\mathcal{M}_k)\supset
\mathcal{N}_{\mathrm{aux}}\supset T^{\ast}\mathcal{M}_k$ with a
fiber over $\phi^{A_k}$:
$\mathcal{F}^{\mathcal{N}_{\mathrm{aux}}}_{\phi^{A}}=
\{(\lambda^{\ast}_{A_k},\phi^{\ast}_{A_k})\}$] by means of a
\emph{BRST-like charge}. This object is nilpotent for $\hbar=0$
with respect to an even Poisson bracket defined in
$C^{\infty}\left(T^{\ast}\mathcal{M}_k\right)$
[$C^{\infty}\left(\mathcal{N}_{\mathrm{aux}}\right)$] and
determines a formal dynamical system subject to first-class
constraints by means of an algorithm which differs from that
of \cite{GMR,GrigorievDamgaard,AlexandrovKontsevichSchwarzZaboronsky}.
To this end, let us consider a functional
$\Omega_k(x_k,\phi^{\ast}_k) \in
C^{\infty}\left(\mathcal{N}_{\mathrm{aux}}\right)$,
$(\varepsilon,\mathrm{gh})\Omega_k=(1,-1)$, constructed
using nilpotent fermionic quantities $V^{\ast}_k$
and $\eta=\mathrm{const}$, $\mathrm{gh}(\eta)=-1$,
\begin{equation}
\Omega_k =V^{\ast}_k S_k(\Gamma_k,\hbar) + \eta \mathcal{S}_0(A)
\equiv \lambda^{\ast}_{A_k}\frac{\delta_l S_k}{\delta
\phi^{\ast}_{A_k}} + \eta \mathcal{S}_0, \label{6}
\end{equation}
as well as using an odd operator $\Delta^k_{\mathrm{d}}$
dual to $\Delta^k$ and an even Poisson bracket $\{\ ,\ \}^k$
nondegenerate in $T^{\ast}\mathcal{M}_k$
\begin{eqnarray}
&& \Delta^k_\mathrm{d} = \eta
(-1)^{\varepsilon_{A_k}}\frac{\delta_l}{\delta
\phi^{A_k}}\frac{\delta_l}{\delta \lambda^\ast_{A_k}}, \ \ \ \{\ ,\
\}^k = \frac{\delta_r \;}{\delta
   x^p_k}\omega^{pq}_{\mathrm{d};k}(x_k)\frac{\delta_l \;}{\delta
   x^q_k},  \label{7} \\
&& \varepsilon(\omega^{pq}_{\mathrm{d};k})=
\varepsilon(\omega^{pq}_{k}) +
1= \varepsilon(x^p_k)+\varepsilon(x^q_k),    \left\|\omega^{
pq}_{\mathrm{d};k}\right\|=\left\|\hspace{-0.5em}\begin{array}{cc}
                             \mathrm{antidiag}\big(%
-\mathbf{{1}}_{n_{k+}},\mathbf{{1}}_{n_{k+}}\big) &\hspace{-0.5em}0 \\
                             0 & \hspace{-0.5em}\mathrm{antidiag}\big(%
\mathbf{{1}}_{n_{k-}},\mathbf{{1}}_{n_{k-}}\big)
                           \end{array}\hspace{-0.5em}\right\|
. \nonumber
\end{eqnarray}
The operator  $V^{\ast}_k$, being a contracting homotopy
with respect to a nilpotent operator ${V}_k$, ${V}_k = \phi^\ast_{A_k}\frac{\delta_l}{\delta
\lambda^\ast_{A_k}}$,
for the  operator   $N_{{V_k}}$, $N_{{V_k}} = [{V}_k,
V^{\ast}_k]_{+}$,
nondegenerate in the fibers
$\mathcal{F}^{\mathcal{N}_{\mathrm{aux}}}_{\phi^{A}}$,
possesses, along with ${V_k}$, the following
properties:\footnote{In general, $\Pi T^{\ast}(T^{\ast}\mathcal{M}_k)$=%
 $\bigl\{(x^p_k, (\phi^{\ast}_{A_k},\lambda^{A_k}))\bigr\}$ admits
 the existence of nilpotent operators $\bigl(U_k, U^{\ast}_k,
 \Pi \Delta^k_{\mathrm{d}}\bigr)$=$\left((-1)^{\varepsilon_{A_k}}\lambda^{A_k}%
 \frac{\delta_l}{\delta \phi^{A_k}}\right.$, $\left.(-1)^{\varepsilon_{A_k}}\phi^{A_k}%
 \frac{\delta_l}{\delta \lambda^{A_k}},
 \eta (-1)^{\varepsilon_{A_k+1}}\frac{\delta_l}{\delta
 \lambda^{A_k}}\frac{\delta_l}{\delta \phi^\ast_{A_k}}\right)$
 similar to $\bigl(V_k, V^{\ast}_k, \Delta^k_{\mathrm{d}}\bigr)$,
 which obey the same properties as in  (\ref{8})
 with the corresponding exchange
 $\bigl(V_k, V^{\ast}_k, \Delta^k_{\mathrm{d}}\bigr) \leftrightarrow \bigl(U_k, U^{\ast}_k,
 \Pi \Delta^k_{\mathrm{d}}\bigr)$; they also anticommute between themselves, $[E, D]_{+}$=0, $E\in
 \bigl\{V_k, V^{\ast}_k, \Delta^k_{\mathrm{d}}\bigr\}, D \in \bigl\{U_k, U^{\ast}_k,
 \Pi \Delta^k_{\mathrm{d}}\bigr\}$, and yield the
 operator  $N_{{U}_k}$=$[U_k, U^{\ast}_k]_{+}$
 nondegenerate in the fibers $\mathcal{F}^{\Pi T^{\ast}(T^{\ast}\mathcal{M}_k)}_{x^p_k}$.}%
 \begin{eqnarray}
    & &  [V^{\ast}_k, \Delta^k]_{+} = 0,\; V^{\ast}_k \left(\mathcal{F},%
   \mathcal{G}\right)^k = \left(V^{\ast}_k \mathcal{F},%
   \mathcal{G}\right)^k - (-1)^{\varepsilon(\mathcal{F})}\left(\mathcal{F},%
  V^{\ast}_k \mathcal{G}\right)^k , \nonumber \\
   & & [V_k, \Delta^k_\mathrm{d}]_{+} = 0,\; V_k \left\{{F},%
   {G}\right\}^k = \left\{V_k {F},%
 {G}\right\}^k + (-1)^{\varepsilon({F})}\left\{{F},%
  V_k {G}\right\}^k . \label{8}
 \end{eqnarray}
 The gauge algebra relations (\ref{1})--(\ref{3})
 and  eqs. (\ref{4}), (\ref{5}) are equivalently
 described using a correspondence between Poisson brackets of opposite
 parities for arbitrary functionals $\mathcal{F}_{\mathrm{t}}(\Gamma_k) \in
 C^{\infty}\left(\Pi T^{\ast}\mathcal{M}_{k}\right)$, $F_{\mathrm{t}%
 }(x_k,\phi^{\ast}_k) = \left[V^{\ast}_k\mathcal{F}_{\mathrm{t}}(\Gamma_k) + \eta%
 \mathcal{F}_{0{}\mathrm{t}}(\phi_k)\right]$, $\mathcal{F}_{0{}\mathrm{t}} \equiv%
 \mathcal{F}_{\mathrm{t}}|_{\mathcal{M}_k}\in \ker
 N_{V_k}$, where $\mathrm{t}=1, 2$,
 \begin{eqnarray}\label{9}
&&    V_k\big\{F_{\mathrm{1}},F_{\mathrm{2}}\big\}^k =
N_{V_k}\big(\mathcal{F}_{\mathrm{1}},
    \mathcal{F}_{\mathrm{2}}\big)^k -\mathcal{F}_{\mathrm{1}}\left(\frac{\overleftarrow{\delta}}{\delta
\phi^{A_k}}\frac{\overrightarrow{\delta}}{\delta \phi^\ast_{A_k}}\big(N_{V_k}-1 \big) -\big(\overleftarrow{N}_{V_k}-1 \big)\frac{\overleftarrow{\delta}}{\delta
\phi^\ast_{A_k}}\frac{\overrightarrow{\delta}}{\delta \phi^{A_k}}\right)\mathcal{F}_{\mathrm{2}},\;
  \\
&& \Delta^k_{\mathrm{d}}F_{\mathrm{t}}= \eta
    \Delta^k\mathcal{F}_{\mathrm{t}};\ \ \
    \left\{\hspace{-0,1em}\Omega_{k},\Omega_{k}\hspace{-0,1em}%
\right\}^{k}\ =\  - V^*_{k}\big(S_{k},\,S_{k}
\big)^{k} - 2 S_{k}\frac{\overleftarrow{\delta}}{\delta
\phi^{A_{k}}}\frac{\overrightarrow{\delta}}{\delta \phi^\ast_{A_{k}}} V^*_{k}S_{k}-2 \eta \mathcal{S}_0,_i\frac{\overrightarrow{\delta}}{\delta A^\ast_{i}}S_{k}   \nonumber,\\
&& \phantom{ \Delta^k_{\mathrm{d}}F_{\mathrm{t}}= \eta
    \Delta^k\mathcal{F}_{\mathrm{t}};\ \ \ } =  - 2 S_{k}\frac{\overleftarrow{\delta}}{\delta
\phi^{A_{k}}}\frac{\overrightarrow{\delta}}{\delta \phi^\ast_{A_{k}}} V^*_{k}S_{k}-2 \eta \mathcal{S}_0,_i\frac{\overrightarrow{\delta}}{\delta A^\ast_{i}}S_{k}, \ \mathrm{for}\  k =\mathrm{min}, \mathrm{ext}  \label{9.1},
\end{eqnarray}
with $\overleftarrow{N}_{V_k}= \frac{\overleftarrow{\delta}}{\delta \phi^\ast_{A_k}}\phi^\ast_{A_k}+\frac{\overleftarrow{\delta}}{\delta \lambda^\ast_{A_k}}\lambda^\ast_{A_k} $ subject to $N_{V_k}\mathcal{F}=  \mathcal{F}\overleftarrow{N}_{V_k} $. From eqs. (\ref{9}), (\ref{10}), it follows that in rank-1 gauge theories
(i.e., $S_{k} = \mathcal{S}_0 + \phi^{\ast}_{A_k}H^{A_k}(\phi_k) \Leftrightarrow V^{\ast}_k S_{k}
\in T^{\ast}\mathcal{M}_{k}, k = \mathrm{ext}$) to whose class one can always reduce an initial gauge model (modulo the conservation
of locality and covariance) the BFV--BRST quantities dual to $S_k, S^{\psi}$ are defined only in $C^{\infty}\left(T^{\ast}\mathcal{M}_{\mathrm{ext}}\right)$. Therefore, only in rank-1 gauge theories does the equation   $\left\{\hspace{-0,1em}\Omega_{k},\Omega_{k}\hspace{-0,1em}%
\right\}^{k} =0$ hold true,  when $\Delta^k S_k =0$,  with allowance for (\ref{1}) and $\big(N_{V_k}-1 \big)S_k=0$.
As a consequence,
 \begin{eqnarray}
&&  {1}/{2}\left\{\hspace{-0,1em}
E_k(x_k),
E_k(x_k)\hspace{-0,1em}\right\}^k = - {1}/{2}V^*_k\left(\hspace{-0,1em}
\mathcal{E}_k(\Gamma_k),
\mathcal{E}_k(\Gamma_k)\hspace{-0,1em}\right)^k = 0, \ \   i \hbar
\Delta^k_{\mathrm{d}} E_k = i \hbar \eta
\Delta^k \mathcal{E}_k = 0, \,\label{10}\\
&& \ \mathrm{with} \  \;E_k \in
\left\{\hspace{-0,1em}\Omega_k,
\Omega^{\psi}\hspace{-0,1em}\right\}, \ \mathcal{E}_k \in
\left\{\hspace{-0,1em}S_k,
S^{\psi}\hspace{-0,1em}\right\},\ \  \Omega^{\psi}
=  \exp\left[\left\{F,
\; \right\}\right]\Omega_k, \label{10.1}
\end{eqnarray}
  with a \emph{gauge boson} $F(\phi_k)$,  $F(\phi_k) = \eta
\psi(\phi_k)$, so that $\Omega^{\psi}
= \Omega_k +  \eta \big(\psi, \, S_k\big)$ for $k=\mathrm{ext}$.\footnote{A
more natural construction of the BV--BFV duality is realized by
a $\theta$-local superfield model with superfields
$\Lambda^{\ast}_A(\theta)$, $\Lambda^{\ast}_A(\theta) =
\lambda^{\ast}_A + \theta \phi^{\ast}_{A}$, instead of
$\Phi^{\ast}_{A}(\theta)$, $\Phi^{\ast}_{A}(\theta) =
\phi^{\ast}_{A}-\theta J_A$, used in \cite{R0,R1,GMR, LMR}.}

\noindent
\paragraph{3. Poisson brackets and triplectic-like algebra of $\Delta^{\mathcal{M}},
{V}^{\mathcal{M}}$ $\left(\mathcal{V}^{\ast},
\mathcal{U}^{\ast}\right)$} Leaving aside the implementation of
a gauge model within the BV method, let us consider a Poisson
supermanifold $\left(\mathcal{M}_0, \{\cdot ,\cdot\}_0\right),
\mathcal{M}_0$ = $\{x^p\}$, $\dim \mathcal{M}_0 = \dim
T^{\ast}\mathcal{M}_{\mathrm{ext}}$,\footnote{In the
infinite-dimensional case, the concept of dimension has to be
clarified. For a vector bundle
$\mathcal{M}_{0}\rightarrow\widetilde{\mathcal{M}}$ with a base
superspace-time $\widetilde{\mathcal{M}}$, it\thinspace
is\thinspace formally\thinspace understood\thinspace
that\thinspace $\dim\mathcal{M}_{0}$\thinspace is\thinspace
the\thinspace dimension\thinspace of\thinspace the\thinspace
fiber\thinspace $\mathcal{F}_{p}^{\mathcal{M}_{0}}$\thinspace
over\thinspace an\thinspace arbitrary\thinspace
$p\in\widetilde{\mathcal{M}}$.} equipped with an even
Poisson bracket defined by a tensor (bivector) field
$\omega^{pq}(x)$ over $\mathcal{M}_0$,
$\omega^{pq}=-(-1)^{\varepsilon_p
\varepsilon_q}\omega^{qp}$, $(\varepsilon, \mathrm{gh})
\omega^{pq} = \bigl(\varepsilon_p +
\varepsilon_q,\mathrm{gh}(x^p)+ \mathrm{gh}(x^q)+2\bigr),
\varepsilon(x^p)=\varepsilon_p$, whose Jacobi identity is
\begin{equation}
    \omega^{qp}\left({\delta_l \omega^{rs}}/{\delta
x^p}\right)(-1)^{\varepsilon_q\varepsilon_s} +
\mathrm{cycle}(q,r,s) = 0,\label{11}
\end{equation}
and also equipped with a covariant derivative $\nabla_p$ in $\mathcal{M}_0$
transforming $\mathcal{M}_0$ into a Poisson supermanifold
with a symmetric connection $\Gamma^p{}_{rs}(x)$,
$\Gamma^p{}_{rs}=(-1)^{\varepsilon_r\varepsilon_s}
\Gamma^p{}_{sr}$ (for a nondegenerate $\omega^{pq}(x)$
providing the existence of quantities $\omega_{pq}(x),
\omega_{pq}=-(-1)^{\varepsilon_p \varepsilon_q}\omega_{qp}$,
$\omega^{pq}\omega_{qr}(-1)^{\varepsilon_q}=\delta^p_r$, $\mathcal{M}_0$
transforms into a Fedosov supermanifold \cite{GL1})
\begin{equation}
    \overleftarrow{\nabla}_r\omega^{pq} =  {1}/{2}\left[({\delta_r
\omega^{pq}})/({\delta x^r}) + 2
\omega^{ps}\Gamma^q{}_{rs}(-1)^{\varepsilon_s(\varepsilon_q+1)}\right]
- (-1)^{\varepsilon_p\varepsilon_q}(p \leftrightarrow q) = 0
.\label{12}
\end{equation}
Let us introduce a manifold $\mathcal{M} = \{(x^p,\eta_p)\}$
locally implemented as an odd (co)tangent bundle over
$\mathcal{M}_0$, $\mathcal{M} = \Pi T^{\ast}\mathcal{M}_0\simeq
\Pi T\mathcal{M}_0$, whose fibers are parameterized by covariantly
constant vectors $\eta_p$, $(\varepsilon, \mathrm{gh})\eta_p =
(\varepsilon_p + 1,-1-\mathrm{gh}(x^p))$, being the antifields for
$x^p$ [$\mathcal{M}_0 = T^{\ast}\mathcal{M}_{\mathrm{ext}}$,
$(x^p;\eta_p) =
(\phi^{A_k},\lambda^{\ast}_{A_k};\phi^{{\ast}}_{A_k},\lambda^{A_k})$].
We next define a functional $T(x,\eta)$ being a scalar w.r.t.
a covariant derivative (extended to act in $\mathcal{M}$)
$\overleftarrow{{\nabla}}_p^{\mathcal{M}}$,
\begin{equation} \label{13}
    T=\frac{1}{2} \eta_p\omega^{pq}(x)\eta_q,\ \ (\varepsilon, \mathrm{gh})T
    = (0,0),\ \  {\overleftarrow{\nabla}}_p^{\mathcal{M}} T=
    \frac{\delta_r T}{\delta x^p} + \frac{\delta_r T}{\delta
    \eta_q}\eta_r \Gamma^r{}_{qp} = 0,
\end{equation}
by virtue of eqs. (\ref{12}) and the relations
${\overleftarrow{\nabla}}_p^{\mathcal{M}} \eta_q = 0$.
Equipping $\mathcal{M}_0$ with a bosonic scalar
density $\rho(x)$ is sufficient to determine a set of covariant
operations characteristic for the supersymplectic manifold
$\mathcal{M}$: an antibracket $(\cdot,\cdot)^{\mathcal{M}}$ and
operators $\Delta^{\mathcal{M}}$, $V^{\mathcal{M}}$,
\begin{eqnarray}
&& (\cdot,\cdot)^{\mathcal{M}} =
\left({\overleftarrow{\nabla}}_p^{\mathcal{M}}
\cdot\right)\frac{\delta_l \cdot }{\delta \eta_p} -
\left(\frac{\delta_r \cdot }{\delta
\eta_p}\right){\overrightarrow{\nabla}}_p^{\mathcal{M}}\cdot
,\;\Delta^{\mathcal{M}} = -(-1)^{\varepsilon_p}\frac{\delta_r \;
}{\delta\eta_p}\left({\overleftarrow{\nabla}}_p^{\mathcal{M}} +
\frac{1}{2}\frac{\delta_r \rho}{\delta
x^p}\right), \label{14}\\
&& V^{\mathcal{M}} = (T, \ )^{\mathcal{M}} = -
\eta_p\omega^{pq}{\overrightarrow{\nabla}}_q^{\mathcal{M}},
\label{15}
\end{eqnarray}
which (using an explicit verification for scalars in ${\mathcal{M}}$)
can be shown to obey the relations of a \emph{triplectic-like
algebra} \cite{GLN}, $[E_1,E_2]_+$ =0 for $E_{1},E_{2} \in
\{\Delta^{\mathcal{M}}$, $V^{\mathcal{M}}\}$, which are consistent
with a Leibniz rule similar to (\ref{8}) for differentiating
an antibracket by any $E_{1},E_{2}$ only in the case of a \emph{flat
Poisson manifold} $\mathcal{M}_0$.\footnote{Namely, for a
vanishing curvature tensor $R^q{}_{prs}(x)$ defined for arbitrary
vectors $T^r(x)$ as follows:
$[{\overleftarrow{\nabla}}_q,{\overleftarrow{\nabla}}_p]T^r(x) = -
(-1)^{\varepsilon_s(\varepsilon_r+1)}T^s(x) R^r{}_{sqp}(x)$.} An
obvious representation of $\mathcal{M}$ as $\Pi T\mathcal{M}_0$
allows one to lift the nondegenerate Poisson structure $\{\cdot,
\cdot\}_0$ to a flat Fedosov manifold $(\mathcal{M}, \{\cdot,
\cdot\})$ by the relation
\begin{equation}
\{ \cdot ,\cdot\} = \left({\overleftarrow{\nabla}}_p^{\mathcal{M}}
\cdot
\right)\omega^{pq}\left({\overrightarrow{\nabla}}_q^{\mathcal{M}}
\cdot \right) + \alpha\frac{\delta_r \cdot }{\delta \eta_p}
\omega_{pq}(-1)^{\varepsilon_p}\frac{\delta_l \cdot }{\delta \eta_q},\;
\alpha=\mathrm{const} \in \mathbb{R}.\label{16}
\end{equation}
A covariant definition for a nilpotent operator
$U^{\mathcal{M}}$ satisfying the relations $[E_1,E_2]_+$ =0
for $E_{1},E_{2} \in \{\Delta^{\mathcal{M}}$, $V^{\mathcal{M}}$,
$U^{\mathcal{M}}\}$ as in Sec. 2 (see footnote 7), in comparison
with the implementation \cite{GLN} of a modified triplectic
algebra, is impossible in terms of an anti-Hamiltonian vector
field, but is provided by equipping $\mathcal{M}_0$ with
an additional Riemann-type nondegenerate even structure
$g_{pq}(x)$,\thinspace $g_{pq}$=$(-1)^{\varepsilon_p
\varepsilon_q}g_{qp}$,\thinspace in\thinspace the\thinspace form
\begin{equation}\label{17}
U^{\mathcal{M}} =  -
\eta_p\omega^{ps}g_{st}\omega^{tq}(-1)^{\varepsilon_s}{\overrightarrow{\nabla}}_q^{\mathcal{M}},\
\left(U^{\mathcal{M}}\right){}^2=0 \Leftrightarrow
{\overrightarrow{\nabla}}_q g_{ps} = 0.
\end{equation}
Since the action of $\nabla$ is undefined as a tensor operation
in the coordinates $x^p$, an explicit definition for an operator
$V^{\ast}{}^{\mathcal{M}}$ ($U^{\ast}{}^{\mathcal{M}}$) of
contracting homotopy w.r.t. $V^{\mathcal{M}}$ ($U^{\mathcal{M}}$)
for  a nondegenerate (as applied to nonconstant functions
in $C^{\infty}\left(\mathcal{M}\right)$) operator $N^{\mathcal{M}}$
($N^{\mathcal{M}}_{U}$) is possible only for a symplectic
$\mathcal{M}_0$, i.e., for $\Gamma^p{}_{rs}$=0. To this end, we
consider the adjoint action of a fermionic functional
$\Omega_T(x,\eta)$, $\Omega_T = - \eta_p x^p$, w.r.t. the non-tensor
bracket (\ref{16}), and then, in order to choose some
operators $\mathcal{V}^{\ast}, \mathcal{U}^{\ast}$ analogous to
$V^{\ast}_k, U^{\ast}_k$ of Sec. 2,  we define an operator
$U^{\ast}{}^{\mathcal{M}}$ using the non-tensor bracket (\ref{14})
with a bosonic $T^{\ast}(x,\eta)$, $T^{\ast} = - (1/2)
x^pg_{pq}(x)x^q(-1)^{\varepsilon_q}$,
  \begin{eqnarray} \label{18}
  &&
V^{\mathcal{M}}+\alpha V^{\ast}{}^{\mathcal{M}}\ = \ \{ \Omega_T, \ 
\} = -\eta_p\omega^{pq}\frac{\delta_l}{\delta x^q}  - \alpha
(-1)^{\varepsilon_p}x^p\omega_{pr} \frac{\delta_l}{\delta \eta_r} , \\
&&  U^{\ast}{}^{\mathcal{M}}\
= \ (T^{\ast}, \ )^{\mathcal{M}}
 =-x^p\left(g_{pq}(x)(-1)^{\varepsilon_q} + \frac {1}{2}x^s\frac{\delta_r
g_{sp}}{\delta x^q}
\right)\frac{\delta_l}{\delta\eta_q},\label{18.1}
\end{eqnarray}
Necessary conditions for the quantities $V^{\ast}{}^{\mathcal{M}}
= 1/\alpha\left[\{ \Omega_T, \, \} - (T, \
)^{\mathcal{M}}\right]$ = $ -(-1)^{\varepsilon_p}x^p\omega_{pr} \big(\delta_l / {\delta\eta_r}\big)$, $U^{\ast}{}^{\mathcal{M}}$ to be nilpotent,
as well as to anticommute between themselves and also
with $\Delta^{\mathcal{M}}$, are given by the fulfilment of
$\left(D,D\right){}^{\mathcal{M}} = 0, \Delta^{\mathcal{M}}D=0$
for $D \in \{T, T^{\ast}\}$, whereas $N^{\mathcal{M}}$ is defined
according to
\begin{eqnarray}\label{19}
 N^{\mathcal{M}} & = & \left[V^{\mathcal{M}},\,V^{\ast}{}^{\mathcal{M}}\right]_+ \ = \ \left[\eta_p\omega^{pq}\frac{\delta_l}{\delta x^q} ,\,(-1)^{\varepsilon_s}x^s\omega_{sr}\frac{\delta_l}{\delta\eta_r} \right]_+  = \\
 &=&
  x^p \frac{\delta_l}{\delta x^p}
+\eta_p
\left(\delta^p_q + \omega^{pr}x^s \frac{\delta_r \omega_{sq}}{\delta x^r} (-1)^{\varepsilon_r(\varepsilon_q+1)+\varepsilon_s} \right)\frac{\delta_l}{\delta\eta_q}. \nonumber
\end{eqnarray}
We next introduce some mutually anticommuting nilpotent operators
$\mathcal{V}^{\ast}$ and $\mathcal{U}^{\ast}$ whose difference
coincides with $- V^{\ast{}\mathcal{M}}$,
\begin{eqnarray}
(\mathcal{V}^{\ast}, \mathcal{U}^{\ast})   =
({1}/{2})\Big(\left[({1}/{\alpha})\left( (T, \ )^{\mathcal{M}}-
\{ \Omega_T, \ \} \right) - (T^{\ast}, \ )^{\mathcal{M}}\right],
 \left[({1}/{\alpha})\left(
\{ \Omega_T, \ \} - (T, \ )^{\mathcal{M}}\right) - (T^{\ast}, \
)^{\mathcal{M}}\right]\Big). \label{20}
\end{eqnarray}
The decomposition (\ref{20}) and the equations for $T, T^{\ast}$ impose
a number of restrictions on the choice of $(g_{pq}$, $\omega^{pq})(x)$
and reduce, by virtue of (\ref{18}), (\ref{18.1}) the quantities $\mathcal{V}^{\ast},
\mathcal{U}^{\ast}$ to a simpler representation. In the particular case
of Darboux coordinates, also assuming the structure of $g_{pq}(x)$ and
the vanishing of the connection $\Gamma^p{}_{rs}(x)$ in $\mathcal{M}_0$,
\begin{eqnarray}
 \left(x^p, \eta_p, \Gamma^p{}_{rs}(x),
\rho(x)\right)& = & \left((\phi^A, \lambda^{\ast}_A), (\phi^{\ast}_A,
\lambda^A),
 0, 1\right), \nonumber \\
 \left[\omega^{pq}, g_{pq}\right](x) & = &  \mathrm{antidiag}\left[\big(- (-1)^{\varepsilon_{A}}\delta^A_B, \delta^A_B\big), \big(%
\delta^A_B, (-1)^{\varepsilon_{A}}\delta^A_B\big)\right],\label{21}
\end{eqnarray}
we obtain the correspondence
\begin{equation}\label{21.1}
\left(V^{\mathcal{M}},
\mathcal{V}^{\ast}, \mathcal{U}^{\ast},
\Delta^{\mathcal{M}}\right) = \left(V_k-U_k, {V}^{\ast}_k,
{U}^{\ast}_k, \Delta^{k} +
(-1)^{\varepsilon_{A}+1}\frac{\delta_l}{\delta\lambda^A}
\frac{\delta_l}{\delta\lambda^{\ast}_A}\right),
\end{equation}
 for $k=\mathrm{ext}$, with account taken of $\omega_{pq}= \mathrm{antidiag}\big((-1)^{\varepsilon_{A}}\delta^A_B, -\delta^A_B\big)$,
 thereby providing $\omega^{pq}\omega_{qr}(-1)^{\varepsilon_q}=\delta^p_r$.

\noindent
\paragraph{4. Quantization rules}
Let us define a vacuum functional and a generating functional
of Green's functions in general coordinates $z^P = (x^p, \eta_p)$,%
\begin{eqnarray}
&& Z_{X}^{\mathcal{M}}=\int d \tilde{z} \mathcal{D}_0(\tilde{x})
 q^{\mathcal{M}}%
(\tilde{z})\exp\left\{  \left[\frac{i}{\hbar}\left(  W + X%
\right)-H \right](\tilde{z},\hbar)\right\},   \label{22} \\%
&& \hspace{-2,5em} Z^{\mathcal{M}}\hspace{-0,05em}\left[J_{\mathcal{V}},J^{{\mathcal{V}}^{\ast}},%
\eta\right]\hspace{-0,1em} = \hspace{-0,1em}\int \hspace{-0,2em}d
\tilde{z}
\mathcal{D}_0(\tilde{x})q^{\mathcal{M}}%
(\tilde{z})\exp\left\{ \hspace{-0,05em}
\frac{i}{\hbar}\left[\hspace{-0,05em}
W\hspace{-0,05em}\left(\tilde{z},\hbar\right)\hspace{-0,05em} +
\hspace{-0,05em}[X+\hspace{-0,05em}i\hbar H]\left(\tilde{x},
\tilde{\eta}- \eta, \hbar%
\right) \hspace{-0,05em} +
\hspace{-0,05em}J^p_{\mathcal{V}}\tilde{\eta}_p \hspace{-0,05em}+
\hspace{-0,05em}
J^{\mathcal{V}^{\ast}}_p\tilde{x}^p\hspace{-0,05em}\right]\hspace{-0,05em}\right\}\hspace{-0,15em}
, \label{23}
\end{eqnarray}
where $J^{\mathcal{V}^{\ast}}_p, {J}^p_{\mathcal{V}}$ form
a redundant set of sources for the corresponding variables $x^p,
\eta_p$ with the properties $(\varepsilon,
\mathrm{gh})J^{\mathcal{V}^{\ast}}_p$ = $(\varepsilon_p, -
\mathrm{gh}(x^p))$ = $(\varepsilon, -\mathrm{gh})J^p_{\mathcal{V}}
+ (1,-1)$; $H, W, X$ denote a set of bosonic functionals providing
a correct reduction, e.g. for $Z_{X}^{\mathcal{M}}$, to the BV partition
function \cite{BV1}, as well as a respective quantum action and
a gauge-fixing bosonic action for irreducible hypergauges $G_a(z)$
satisfying generalized master-equations and one more restriction
on $W$,
\begin{equation}
\Delta^{\mathcal{M}} \exp
\left[\left({i}/{\hbar}\right)E(z,\hbar)\right] = 0,\ E \in
\{W,X+i\hbar H\}, \ \mathcal{V} W(z,\hbar) = 0 ,
\label{24}%
\end{equation}
for the first of which a proper solution in $\mathcal{M}$ is provided
only by $X$, while $W(z,\hbar)$ is subject to a boundary condition
$W(z,\hbar)_{|\eta = \hbar =0} = \mathcal{S}_0(x)$, with the classical
action $\mathcal{S}_0$, and also subject to $\mathcal{U}^{\ast}W(z,\hbar) = 0$,
for a non-Fedosov (merely symplectic) $\mathcal{M}_0$.
At the same time, the density function $\mathcal{D}_0({x})$, which
determines an invariant measure in $\mathcal{M}$ (corresponding
for $\alpha = 1$ to the even bracket (\ref{16}), so that the functional
$\rho$ (\ref{14}) is defined as $\rho =
\ln\;\mathrm{sdet}^{-1}\|\omega^{pq}({x})\|$, following \cite{GLN}),
and the weight functional $q^{\mathcal{M}%
}[z]$ are given by%
\begin{equation}
\mathcal{D}_0({x}) =
\mathrm{sdet}^{-1}\|\omega^{pq}({x})\|,\;q^{\mathcal{M}}(z)=\delta\left(G_{a_{1}%
}^{\mathcal{V}^{\ast}}(z)\right)
,\,a_{1}=1,\ldots,\,1/2\left(\dim_+\mathcal{M}_0 +\dim_-
\mathcal{M}_0\right). \label{25}%
\end{equation}
In (\ref{23}), (\ref{25}), according to the decomposition of the operator
$V^{\ast\mathcal{M}}$ in (\ref{20}), we have used a polarization
of $V^{\mathcal{M}}$ into the corresponding difference of operators,
\begin{eqnarray}
&& V^{\mathcal{M}} = \mathcal{V} -  \mathcal{U}:
\left(\mathcal{V}, \mathcal{U}\right) =
\frac{1}{2}\eta_p\left(-\omega^{pq} +
  \omega^{ps}g_{st}\omega^{tq}(-1)^{\varepsilon_s}, \omega^{pq} +
  \omega^{ps}g_{st}\omega^{tq}(-1)^{\varepsilon_s}\right){\overrightarrow{
  \nabla}}_q^{\mathcal{M}}, \label{26}
\end{eqnarray}
with a \thinspace nilpotent  anticommuting\thinspace $\mathcal{V},
\mathcal{U}$.\thinspace In\thinspace turn,\thinspace the\thinspace
independent\thinspace functions\thinspace
$G_{a_{1}}^{\mathcal{V}^{\ast}}(z)$=0, playing the role
of \emph{second-level hypergauge conditions} in the formalism
\cite{BatalinTyutin1}, are necessary to retain the
explicitly  cova\-ri\-ant form\thinspace of\thinspace
both\thinspace functionals\thinspace
$Z_{X}^{\mathcal{M}}$,\thinspace $Z^{\mathcal{M}}$.\thinspace
The\thinspace independent\thinspace functions\thinspace
 $G_{a_{1}}^{\mathcal{V}^{\ast}}%
(z)$\thinspace are\thinspace equivalent\thinspace to \thinspace
a \thinspace (known explicitly only for a \thinspace
symplectic $\mathcal{M}_0$)\thinspace set\thinspace
of\thinspace functions\thinspace $\mathcal{V}^{\ast}\eta_{p}$,
$G_{a_{1}}^{\mathcal{V}^{\ast}}%
(z) =\left[ Y_{a_{1}}^p(z)\mathcal{V}^{\ast}\eta_{p}\right]$
with\thinspace a certain\thinspace  $Y_{a_{1}}^p(z)$,\thinspace
so\thinspace that
\begin{equation}
\mathrm{rank}\Bigl\|  {\overrightarrow{
  \nabla}}_p^{\mathcal{M}} E_{\mathrm{t}}(z%
)\Bigr\|  _{{\overrightarrow{
  \nabla}}^{\mathcal{M}}  W= \delta W /\delta \eta = {\overrightarrow{
  \nabla}}^{\mathcal{M}}X =
  \delta X /\delta \eta = G^{\mathcal{V}^{\ast}}=0}= \frac{1}{2}\dim \mathcal{M}_0,
\,(E_{1},E_{2})=(G_{a_{1}}^{\mathcal{V}^{\ast}},\mathcal{V}^{\ast}\eta_{p}),
\label{27}%
\end{equation}
 and define the functional $H(z,\hbar)$ in (\ref{22})--(\ref{24})  in an explicitly covariant form \cite{BatalinTyutin1}:
\begin{equation}\label{28}
H(z,\hbar) = - \frac{1}{2}\ln \left\{J(z)
\mathcal{D}^{-1}_0(x)\mathrm{sdet} M  \right\},\ M =
\textstyle\left\|\begin{array}{cc}
\bigl(F^{\mathcal{U}^{\ast}}_{a_1}(z),G_{\mathcal{V}}^{c_1}%
(z)\bigr)^{\mathcal{M}}   & \bigl(F^{\mathcal{U}^{\ast}}_{a_1}(z),F_{\mathcal{U}}^{d_1}%
(z)\bigr)^{\mathcal{M}} \\
\bigl(G^{\mathcal{V}^{\ast}}_{b_1}(z),G_{\mathcal{V}}^{c_1}%
(z)\bigr)^{\mathcal{M}} & \bigl(G^{\mathcal{V}^{\ast}}_{b_1}(z),F_{\mathcal{U}}^{d_1}%
(z)\bigr)^{\mathcal{M}} \\
\end{array}\right\|,
\end{equation}
where the functions $(F_{\mathcal{U}}^{a_1},
F^{\mathcal{U}^{\ast}}_{b_1},G_{\mathcal{V}}^{c_1})$=%
$(\widetilde{Z}^{a_1}_p{\mathcal{U}}x^p,
Z_{b_1}^p{\mathcal{U}^{\ast}}\eta_p,
\widetilde{Y}^{c_1}_p{\mathcal{V}}x^p)$ with certain
$(\widetilde{Z}^{a_1}_p, Z_{b_1}^p, \widetilde{Y}^{c_1}_p)(z)$
determi\-ne an\thinspace invertible\thinspace change\thinspace
of\thinspace variables, $z^P \rightarrow
\overline{z}{}^P$=$(F^{\mathcal{U}^{\ast}}_{a_1},
G^{\mathcal{V}^{\ast}}_{b_1}, G_{\mathcal{V}}^{c_1},
F_{\mathcal{U}}^{d_1})$,\thinspace with\thinspace the\thinspace Jacobian\thinspace $J$=%
$\mathrm{sdet}\left\|\delta\overline{z}{}^Q /\delta
z^P\hspace{-0.1em}\right\|$.

The basic properties of the functionals $Z_{X}^{\mathcal{M}}$,
$Z^{\mathcal{M}}$ are encoded by a \emph{generalized generator}
$s^{\mathcal{M}}$ of \emph{BRST-like transformations} with an
arbitrary bosonic functional  $R^{\mathcal{M}}(z)$,%
\begin{equation}
s^{\mathcal{M}} = \left({\hbar}/{i}\right)T^{-1}(z)\left(\ \
\;,T(z)R^{\mathcal{M}}(z)\right) ^{\mathcal{M}},\ T(z)=\exp\left[
\left({i}/{\hbar}\right)\left( W-X-i\hbar H\right)(z)
\right]  . \label{29}%
\end{equation}
For instance, the BRST transformations with a constant $\mu$,
$\delta_{\mu}z^{P} = s^{\mathcal{M}} z^{P}\mu$, for
$Z_{X}^{\mathcal{M}}$ and $Z^{\mathcal{M}}\left[0, 0, \eta\right]$,
are derived using (\ref{29}), with $R^{\mathcal{M}}$=1, and
also using additional equations providing the BRST invariance
of $q^{\mathcal{M}}$, namely,
\begin{equation}
\left(  G_{a_{1}}^{\mathcal{V}^{\ast}}(z),T(z)\right){}
^{\mathcal{M}}=0\Longleftrightarrow\delta_{\mu
}G_{a_{1}}^{\mathcal{V}^{\ast}}(z)=0. \label{30}%
\end{equation}
Thus the constant (being covariantly constant for a flat Fedosov
$\mathcal{M}_0$) functions $Y_{a_{1}}^p(x)$ belong to solutions
of eqs. (\ref{30}), imposing strong restrictions on the geometry
of $\mathcal{M}$.

The Ward identity for the functional $Z^{\mathcal{M}}$ and
the gauge-independence of the S-matrix follow from
the master-equations (\ref{24}), the transformation (\ref{29}),
and also from some additional equations for $G_{a_{1}}^{\mathcal{V}^{\ast}}(z)$ (\ref{30}),
along the lines of \cite{R1,GMR}. The transformation (\ref{29})
in $Z_{X+\delta X}^{\mathcal{M}}$, with the choice of
$R^{\mathcal{M}}$ as $R^{\mathcal{M}}\mu$=${\delta} Y(z)$, where
the fermionic functional ${\delta} Y(z)$ is found from the
equality $\delta X(z)$ = $Q(X,H){\delta}Y(z)$ [with a nilpotent
$Q(X,H)$, $Q(X,H) = (X+i\hbar H,\ )^{\mathcal{M}} - i\hbar
\Delta^{\mathcal{M}}$], being a general non-vanishing solution
for the linearized equation $Q(X,H)\delta X(z)$=0, establishes
that $Z_{X+\delta X}^{\mathcal{M}}$=$Z_{X}^{\mathcal{M}}$. In
turn, after exponentiating the quantity $q^{\mathcal{M}}$
in the functional integral (\ref{23}),
\begin{equation*}
q^{\mathcal{M}}(\tilde{z}) =\int
d{}\tilde{\lambda}_{(2)}
\exp\bigl\{\bigl(i/\hbar\bigr)G_{a_{1}}^{\mathcal{V}^{\ast}}(\tilde{z})
\tilde{\lambda}^{a_{1}}_{(2)}\bigr\},\
\varepsilon(\tilde{\lambda}^{a_{1}}_{(2)}) = \varepsilon(G_{a_{1}}^{\mathcal{V}^{\ast}})%
=\varepsilon_{a_1},\, \mathrm{for} \ \delta_{\mu
}\tilde{\lambda}^{a_{1}}_{(2)} =0,
\end{equation*}
 the corresponding Ward identity (with a loss of covariance, ${\overrightarrow{
  \nabla}}_q^{\mathcal{M}} \to \big(\delta_l / \delta {x}^p\big)$) is given by
\begin{eqnarray}
 \hspace{-0.7em} &\hspace{-0.7em}&\hspace{-0.7em} \left[J^{\mathcal{V}^{\ast}}_p +  \hspace{-0.1em} \left\{\hspace{-0.1em}\frac{\delta_r W}{\delta \tilde{x}{}^p} + (-1)^{\varepsilon_{a_1}}
  \langle\tilde{\lambda}^{a_{1}}_{(2)}\rangle \frac{\delta_r G_{a_{1}}^{\mathcal{V}^{\ast}}}{\delta \tilde{x}{%
  }^p} +\frac{\hbar}{i}\mathcal{D}^{-1}_0\frac{\delta_r \mathcal{D}_0}{\delta \tilde{x}{%
  }^p}\hspace{-0.1em}\right\}
  \hspace{-0.1em}\left(\hspace{-0.1em}\frac{\hbar}{i}\frac{\delta_l}{\delta J^{\mathcal{V}^{\ast}}},
  \frac{\hbar}{i}\frac{\delta_l}{\delta J_{\mathcal{V}}}\hspace{-0.1em}\right)
  \hspace{-0.1em}\right]\frac{\delta_l}{\delta \eta_p}
  Z^{\mathcal{M}}\Big[J_{\mathcal{V}}, J^{{\mathcal{V}}^{\ast}},
\eta\Big]\hspace{-0.1em} =0.  \label{31}
\end{eqnarray}
To obtain eq. (\ref{31}), we start with a functional averaging
of the master-equation (\ref{24}) for $(X+i\hbar H)(\tilde{x}$,
$\tilde{\eta} - \eta)$ w.r.t. the weight functional
$\exp\bigl\{\left(\hspace{-0.05em}i/\hbar\hspace{-0.05em}\right)\bigl[W
\left(\hspace{-0.05em}\tilde{z}\hspace{-0.05em}\right) +
G_{a_{1}}^{\mathcal{V}^{\ast}}(\tilde{z})
\tilde{\lambda}^{a_{1}}_{(2)} + J^p_{\mathcal{V}}\tilde{\eta}_p +
J^{\mathcal{V}^{\ast}}_p\tilde{x}^p\bigr]\bigr\}$, and then
use a non-tensor character of $x^p$ to integrate by parts
with allowance for $({\delta_l}/{\delta \tilde{\eta}_p} +
{\delta_l}/{\delta \eta_p})(X+i\hbar H)(\tilde{\eta}-\eta) = 0$. Note that
the symbol $\langle F(\tilde{z},\tilde{\lambda}_{(2)})\rangle$ denotes
a source-dependent average expectation value for
a quantity $F(\tilde{z},\tilde{\lambda}_{(2)})$ corresponding to a gauge-fixing $X$
w.r.t.  $Z^{\mathcal{M}}\hspace{-0,05em}\left[J_{\mathcal{V}},J^{{\mathcal{V}}^{\ast}},%
\eta\right]$ in the presence of an external $\eta_p$:
\begin{eqnarray}
&& \left\langle F \right\rangle _{X,J_{\mathcal{V}},J^{{\mathcal{V}}^{\ast}}} = {Z^{\mathcal{M}}}^{-1}(J_{\mathcal{V}},J^{{\mathcal{V}}^{\ast}})\int \hspace{-0,2em}d
\tilde{z}\,d{}\tilde{\lambda}_{(2)}
\mathcal{D}_0(\tilde{x})%
\exp\left\{ \hspace{-0,05em}
\frac{i}{\hbar}\left[\hspace{-0,05em}
W\hspace{-0,05em}\left(\tilde{z},\hbar\right)\hspace{-0,05em} +
\hspace{-0,05em}[X+\hspace{-0,05em}i\hbar H]\left(\tilde{x},
\tilde{\eta}- \eta, \hbar%
\right) \hspace{-0,05em}\right.\right. \label{aexv} \\
&& \left.\left.\phantom{\left\langle F \right\rangle _{X,J_{\mathcal{V}},J^{{\mathcal{V}}^{\ast}}} =}  + G_{a_{1}}^{\mathcal{V}^{\ast}}(\tilde{z})
\tilde{\lambda}^{a_{1}}_{(2)} +
\hspace{-0,05em}J_{\mathcal{V}}\tilde{\eta} \hspace{-0,05em}+
\hspace{-0,05em}
J^{\mathcal{V}^{\ast}}\tilde{x}\hspace{-0,05em}\right]\hspace{-0,05em}\right\}\hspace{-0,15em},\ \mathrm{with} \ \left\langle 1\right\rangle _{X,J_{\mathcal{V}},J^{{\mathcal{V}}^{\ast}}}=1.  \nonumber %
\end{eqnarray}

Having chosen the Darboux coordinates for $z^P$, along with
the conditions (\ref{21}), supplemented by the relations $\left(Z_{a_1}^p,
{Y}_{a_1}^p, \widetilde{Z}^{a_1}_p, \widetilde{Y}^{a_1}_p\right)=
\left(\delta^{Ap},\delta_A^p,\delta^A_{p},\delta_{Ap}\right)$,
for which the sources can be written as $\left(J^p_{\mathcal{V}};
J^{\mathcal{V}^{\ast}}_p\right)=\left(- I^{\ast A}, I_A ;
J_A,J^{\ast A}\right)$ and $(\mathcal{D}_0,H)=(1,0)$, we obtain
a coincidence between the (independent of $J^{\ast{}A}$)
functional $Z^{\mathcal{M}}[I_{A}, I^{\ast A}, J_A, \phi^{\ast},
\lambda]$ for ${\lambda^A = 0}$ and the generating functional
of Green's functions $\mathcal{Z}(\theta)_{|\theta=0}$ of the
superfield quantization \cite{GMR} under the evident
designations $(\phi^A, \phi^{\ast}_A,\lambda^A)=(\varphi^a,
\varphi^{\ast}_a,\lambda^a)$.

In conclusion, we note that the suggested quantization procedure
allows one to introduce an effective action and can be elaborated
both under reducible hypergauges \cite{R1} and in the multilevel
formalism. Besides, it suggests a possibility for an explicit
correspondence between the BRST and BRSTantiBRST
quantization approaches in the $N=2$ superfield formalism.

\textbf{Acknowledgments} The author is grateful to P.M. Lavrov and
 P.Yu. Moshin
for useful discussions.

\end{document}